\documentclass[9pt,twocolumn,twoside]{opticajnl}
\journal{opticajournal}
\setboolean{shortarticle}{true}

\title{Sub-terahertz optomechanics}

\author[1]{Jiacheng Xie}
\author[1]{Mohan Shen}
\author[1,*]{Hong X. Tang}
\affil[1]{Department of Electrical Engineering, Yale University, New Haven, CT 06511, USA}
\affil[*]{hong.tang@yale.edu}

\begin{abstract}
We demonstrate optomechanics in the sub-terahertz regime. An optical racetrack resonator, patterned from thin-film lithium niobate, is suspended to support mechanical structures oscillating at these extremely high frequencies, which are read out through cavity optomechanical coupling.  Our hybrid platform paves the way for advancing mechanical systems in the quantum regime at elevated temperatures.
\end{abstract}
\setboolean{displaycopyright}{false}

\begin{document}
\maketitle
Extremely high-frequency mechanical resonators in the sub-terahertz (THz) regime are advantageous in quantum mechanics research since they are readily cooled to their ground state even at kelvin (K) temperatures \cite{xie2023sub}. This reduced cooling requirement allows quantum devices to operate outside of expensive and bulky milli-kelvin (mK) dilution refrigerators whose cooling capacity is also limited. For instance, the thermal occupation of a 100\,gigahertz (GHz) resonator is only 0.008 quanta at 1K, resulting in a 99.2\% probability of the resonator being in the ground state – the same as that of a 10\,GHz microwave resonator in a 100\,mK environment. Therefore, frequency scaling to sub-THz frequencies provides a viable pathway for studying scaled quantum structures and implementing quantum phononic circuits at elevated temperatures. On the other hand, by integrating sub-THz mechanics with multi-hundred-THz optics to form a hybrid optomechanical system, we are able to convert millimeter-wave signals into optical signals, establishing the linkage of remote quantum networks through fibers. 

To achieve this task, we resort to newly-emerged thin-film lithium niobate (TFLN). Its wide optical transparency window, large Pockel's coefficient, and second-order nonlinearity, along with excellent piezoelectric properties make it a great candidate for a variety of optical, microwave and mechanical applications, such as frequency doublers \cite{wang2017second,luo2018highly,lu2019periodically,ye2022second}, electro-optic modulators \cite{mercante2018thin,wang2018integrated,li2020lithium, shen2024photonic} and converters \cite{holzgrafe2020cavity,mckenna2020cryogenic, xu2021bidirectional}, piezo-optomechanical converters \cite{jiang2019lithium, shen2020high}, electromechanical filters \cite{yang20194, barrera202338}, etc. 

\begin{figure}[h]
\centering\includegraphics[width = 0.95\linewidth]
{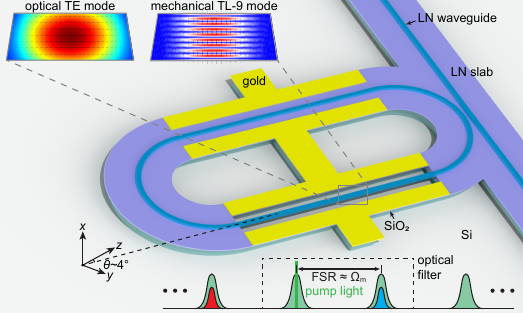}
\caption{Schematic of the sub-THz electro-optomechanical resonator (not to scale). The simulated displacement field for the TL-9 mode of the suspended waveguide is shown in the  top-left inset, along with the modal profile of the optical TE mode. The mechanical resonant frequency matches the optical FSR.}\label{fig1}
\end{figure}

\begin{figure}[t]
\centering\includegraphics[width = 0.95\linewidth]{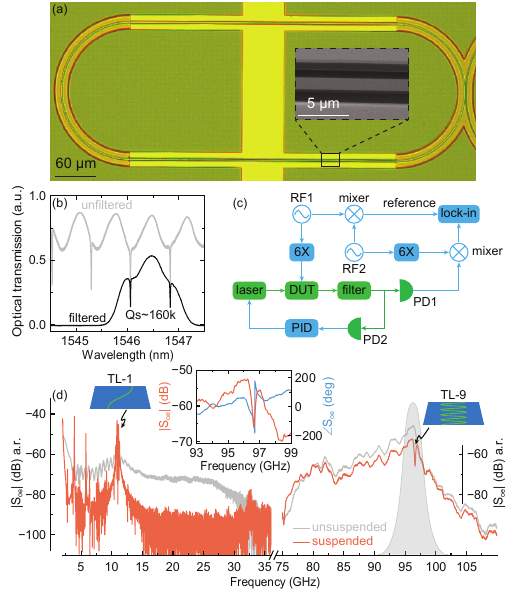}
\caption{(a) Device optical micrograph. The inset shows the SEM image of a suspended waveguide section. (b) Unfiltered and filtered optical transmission. Note that the filter has an insertion loss of around 1-2dB at its center wavelength. (c) W-band $\rm{S_{oe}}$ measurement schematic. A PID loop is employed to lock the laser to an optical resonance. See Ref.\,\cite{xie2023sub} for more information about the construction of W-band VNA. (d) RF-to-optical conversion spectrum $\rm{S_{oe}}$ (arbitrary reference). Inset shows the amplitude and phase response of the TL-9 mode.}
\label{fig2}
\end{figure}

Fig.\,\ref{fig1} shows the design of an electro-optomechanical resonator on $x$-cut TFLN, which integrates an optical racetrack resonator with an electromechanical thickness-longitudinal (TL)-mode resonator. In this configuration, one of the optical resonator's straight waveguide sections is suspended by chemically removing the silicon dioxide beneath the lithium niobate (LN) layer using buffered oxide etchant (BOE). By depositing gold electrodes on the sides of the suspended waveguide, we can efficiently excite its mechanical TL modes with an in-plane electric field, utilizing the $e_{12}$ piezoelectric coupling element \cite{xie2024high}. To target the sub-THz mechanical mode, we design the free spectral range (FSR) of the optical resonator to be $\sim$97\,GHz, which matches the 9-th order TL mode of a 300\,nm-thick LN film and leads to enhanced optomechanical readout efficiency, as illustrated in the bottom-right corner of Fig.\,\ref{fig1}. The simulated mode displacement is shown in the top-left inset of Fig.\,\ref{fig1}, where the color surface represents the displacement amplitude and the black arrows mark the displacement direction. 

Fig.\,\ref{fig2}(a) shows an optical micrograph of the racetrack resonator and a scanning electron microscopy (SEM) image of a suspended waveguide section. Fig.\,\ref{fig2}(b) displays the device's optical transmission before and after applying a 1-nm-bandwidth optical filter. The TE mode FSR is around 96\,GHz. For the measurement, a telecom-wavelength laser (pump) is locked to the optical resonance around 1547\,nm, and microwave signals are routed to the electrodes via an RF probe. At mechanical resonant frequencies, the applied electric field will excite the TL modes and, in turn, modulate light in the suspended optical waveguide via the moving-boundary and elasto-optic effect. The pump and scattered light are subsequently sent to a fast photodetector to produce a beating signal. This beating signal is then demodulated to obtain the RF-to-optical conversion spectrum $\rm{S_{oe}}$. Another mechanism --- the electro-optic (EO) effect --- also contributes to the coherent scattering of the optical pump. This effect, known for its broadband feature, has numerous applications \cite{zhang2022systematic,valdez2023100,juneghani2023integrated} on its own. In our case, however, it arises as an interference to the resonant electro-optomechanical modulation. Fortunately, this effect can be suppressed by aligning the suspended resonator along the crystallographic direction that minimizes the $r_{11}$ electro-optic coefficient, where ``1" denotes the horizontal direction in the cross-sectional view. By applying tensor rotation, we have $r_{11} = r_{yy,y}\cos^3{\theta}-(r_{xx,z}+2r_{yz,y})\sin\theta\cos^2\theta-r_{zz,z}\sin^3\theta$, where $\theta$ represents the clockwise rotational angle on $x$-cut LN film marked by the coordinate system in Fig.\ref{fig1}, and $r_{ij,k}$ represents LN's EO coefficient. In the experiment, we choose a $\theta$ around $4.3^\circ$ to ensure that $r_{11}$ approximates zero.

The measured $\rm{S_{oe}}$ spectrum is shown in Fig.\,\ref{fig2}(d). The lower- and higher-frequency spectra are acquired using a commercial vector network analyzer (VNA) and a home-built W-band VNA (Fig.\,\ref{fig2}(c)), respectively. By separately probing at the suspended and unsuspended sides of the racetrack, we are able to distinguish the mechanical signal. From the $\rm{S_{oe}}$ spectrum, we reveal strong resonances at TL-1 and TL-9 modes located at around 10.9 and 97\,GHz respectively. TL-3 and TL-7 peaks are also discernible at 32.6 and 75\,GHz, respectively, but with significantly smaller signal amplitudes due to the large frequency mismatch with the FSR of the racetrack. The linear relationship between resonant frequency and mode order also indicates thickness-mode characteristics, consistent with the findings in Ref.\,\cite{xie2023sub}. In addition to mechanical modulation, the residual EO conversion peaks around 96\,GHz, which matches the optical FSR (Fig.\,\ref{fig2}(d) grey Lorentzian shape). At room temperatures, the quality factors of these thickness modes are on the order of hundreds, which are expected to increase at cryogenic temperatures \cite{shen2020high}. For the TL-9 mode, the calculated vacuum optomechanical coupling due to the moving-boundary effect $g_\mathrm{om,mb}/(2\pi)$ is approximately $1.1$\,kHz and that due to the elasto-optic effect $g_\mathrm{om,el}/(2\pi)$ is approximately $-0.8$\,kHz. Both decrease inversely proportional to the square root of the mode order, $\sqrt{n}$, to a first-order approximation. Compared to other optomechanical platforms, such as optomechanical crystals, they can achieve hundreds of kHz coupling rates, benefiting from their small mode volumes. However, their compact size poses challenges in efficient electromechanical transduction and in scaling up to sub-THz frequencies. On the other hand, Brillouin scattering platforms offer more flexibility in frequency scaling, but generally have lower optomechanical coupling rates than optomechanical crystals and encounter difficulties with stringent phase-matching conditions \cite{eggleton2019brillouin}.

In conclusion, we devised a sub-THz eletro-optomechanical platform where the mechanical resonant frequency matches the optical FSR. Leveraging this setup, we revealed sub-THz mechanical resonances through optomechanical readout techniques. Sub-THz optomechanics holds promise for fundamental quantum phononics studies and could play an important role in serving as quantum transduction devices that are more immune to thermal excitation.

\begin{backmatter}
\bmsection{Funding} Air Force Office of Sponsored Research (AFOSR MURI FA9550-23-1-0338). Support for lithium niobate thin film preparation is provided by the US Department of Energy Co-design Center for Quantum Advantage (C2QA) under Contract No. DE-SC0012704. 
\bmsection{Acknowledgments} The authors would like to thank Y. Sun, K. Woods, L. McCabe, and M. Rooks for assistance in the device fabrication.
\bmsection{Disclosures} The authors declare no conflicts of interest.
\bmsection{Data Availability} Data are available upon reasonable request.
\end{backmatter}

\bibliography{main}
\end{document}